\documentstyle{article}
\begin{document}

\begin{flushright}
NUP-A-98-7 \\
\end{flushright}
\hspace{\fill}
\vspace{14mm}
\begin{center}
\Large\bf{Anomalous Gauge Theories \\
with Antisymmetric Tensor Fields}
\end{center}
\hspace{\fill}
\begin{center}
{\large 
$\rm{Shinichi \; Deguchi}^{\,\it a}\!\!$
\footnote{E-mail: deguchi@phys.cst.nihon-u.ac.jp}, 
$\rm{Tomoaki \; Mukai}^{\,\it{b}}\!\!$ and 
$\rm{Tadahito \; Nakajima}^{\,\it{b}}\!\!$
\footnote{E-mail: nakajima@phys.cst.nihon-u.ac.jp}} 
\end{center}

\begin{center}
{\it ${}^{a)}$ 
Atomic Energy Research Institute, 
College of Science and Technology, \\
Nihon University, Tokyo 101-8308, Japan} 
\end{center}
\begin{center}
{\it ${}^{b)}$ Department of Physics, 
College of Science and Technology, \\
Nihon University, Tokyo 101-8308, Japan} 
\end{center}

\hspace{\fill}
\vspace{5mm}
\begin{flushleft}
\bf{Abstract}
\end{flushleft}

We derive a dual theory of the four-dimensional anomalous U(1) gauge theory 
with a Wess--Zumino (WZ) term and with a St\"uckelberg type mass term  
by means of a duality transformation at each of the classical and quantum 
levels. It is shown that in the dual anomalous U(1) gauge theory, 
the $BF$ term with a rank-two antisymmetric tensor 
field plays the roles of the WZ term as well as the mass term of the U(1) 
gauge field. Similar anomalous U(1) gauge theory with $BF$ term is considered  
in six-dimensions by introducing a rank-four antisymmetric tensor field.  
In addition to this theory, we propose a six-dimensional anomalous U(1) 
gauge theory including an extended $BF$ term with a rank-two antisymmetric 
tensor field, discussing a difference between the two theories. 
We also consider a four-dimensional anomalous SU(2)$\times$U(1) gauge theory 
with $BF$ term and recognize a crucial role of the $BF$ term in  
cancelling the non-abelian chiral anomaly.  

\vspace{4mm}


\newpage
\section{Introduction}

The consistent quantization of anomalous gauge theories has been studied by 
many authors for about the past 10 years [1-6]. In particular, 
the chiral Schwinger model, an anomalous chiral U(1) gauge theory 
in two dimensions, has been investigated in detail under advantageous 
conditions peculiar to two dimensions. 
A remarkable observation is that because of the chiral anomaly, a hidden 
physical degree of freedom occurs in the theory [1]. 
This is well understood by adding a suitable Wess-Zumino (WZ) term 
to the classical action of the chiral Schwinger model 
so as to restore the gauge symmetry at the quantum level [2].   
The origin of such a WZ term can be assigned to the gauge-volume integration 
in the path-integral quantization [3].

The idea of introducing a WZ term is, of course, applicable to anomalous 
gauge theories in higher dimensions. In fact, suggesting this idea, 
Faddeev and Shatashvili have argued for consistent quantization 
of an anomalous chiral gauge theory in four dimensions [4]. 
Since then, several authors have particularly studied the case of  
the U(1) gauge group on the reasonable assumption 
that the chiral gauge symmetry is spontaneously broken (via the anomaly) 
[5,6]. The symmetry breakdown makes the U(1) gauge field massive  
and enables us to adopt perturbative approaches. 
As an effective theory describing the broken phase, 
we can take the anomalous massive U(1) gauge theory in four dimensions 
whose action includes a WZ term and  
a St\"uckelberg type mass term containing a WZ scalar field [6]. 
This theory is consistently quantizable, though renormalizability will be 
spoiled owing to the WZ term.

Besides the St\"uckelberg formalism and the Higgs model, 
there is an alternative massive gauge theory called the topologically massive 
gauge theory, in which a topological term 
called the $BF$ term functions as mass terms of gauge fields [7,8]. 
The topologically massive abelian gauge theory (TMAGT) in four dimensions 
consists of a U(1) gauge field $A_{\mu}$ and a rank-two antisymmetric 
tensor field $B_{\mu\nu}$ [7]. It is not difficult to show, 
both at the classical and quantum levels,  
that the TMAGT is a dual version of the abelian St\"uckelberg formalism.

Taking this duality into account and noting that 
the anomalous massive U(1) gauge theory involves the abelian 
St\"uckelberg formalism, we can find a theory that is dual to  
the anomalous massive U(1) gauge theory and that involves the TMAGT.  
Such a dual theory will indeed be obtained in the next section 
with a suitable modification of the gauge transformation rule of  
$B_{\mu\nu}$ defined in the TMAGT. 
After the modification, the $BF$ term plays the roles of the WZ term 
as well as the mass term of $A_{\mu}$; 
$B_{\mu\nu}$ contributes importantly to cancelling the chiral anomaly. 
The modified gauge transformation rule agrees with the one found in 
$N=1$ supergravity coupled to $N=1$ supersymmetric Maxwell theory in ten 
dimensions [9].

An anomaly-cancellation mechanism in which a rank-two antisymmetric 
tensor field plays a crucial role has been argued in superstring theories 
[10].  This mechanism, known as the Green--Schwarz mechanism, works 
only if the gauge group is chosen to be SO(32) or 
${\rm E}_{8}\times{\rm E}_{8\,}$. 
A non-abelian version of the modified gauge transformation rule 
is also essential to the Green--Schwarz mechanism.

In this paper we would like to explain that anomaly cancellation due to  
an antisymmetric tensor field occurs not only in superstring theories but also 
in (dual versions of) ordinary anomalous gauge theories.  
We demonstrate that $BF$ terms and their generalizations  
function as WZ terms when suitable gauge transformation rules are imposed 
on the antisymmetric tensor fields.

The present paper is organized as follows. 
Section 2 derives, at the classical level, a dual theory of the 
anomalous massive U(1) gauge theory in four dimensions. 
In Section 3, the duality found in Section 2 is established at the quantum 
level by using the path-integral quantization based on the BRST formalism. 
Section 4 considers two kinds of anomalous U(1) gauge theories in 
six dimensions. One of them has the $BF$ term with a rank-four 
antisymmetric tensor field and is dual to the six-dimensional anomalous 
U(1) gauge theory with a WZ term and with a St\"uckelberg type mass term.  
The other theory has the $BF^{2}$ term, a generalized $BF$ term, 
with a rank-two antisymmetric tensor field.  
In each theory, the six-dimensional chiral anomaly vanishes by  
virtue of the $BF$ or $BF^{2}$ term.  
Section 5 deals with an anomalous non-abelian gauge theory in four 
dimensions with the gauge group SU(2)$\times$U(1). 
The $BF$ term in this theory functions both as the WZ term for  
the non-abelian chiral anomaly and as the mass term of the U(1) gauge field. 
Section 6 is devoted to summary and discussion.

\setcounter{enumi}{\value{equation}}
\addtocounter{enumi}{1}
\renewcommand{\theequation}{\thesection.\theenumi\alph{equation}}
\setcounter{equation}{0}

\renewcommand{\theequation}{\thesection.\arabic{equation}}
\setcounter{equation}{0}

\section{Dual version of an anomalous U(1) gauge theory} 
\setcounter{equation}{0}
\vspace{0.5mm}

Let us begin by discussing an anomalous massive U(1) gauge theory in four 
dimensions [6] that is defined by the lagrangian 
\begin{eqnarray}
\widetilde{\cal L}={\cal L}_{A}+{\cal L}_{\phi}+{\cal L}_{\rm WZ}+
{\cal L}_{\psi}
\end{eqnarray}
%(2.1)
%
with
\begin{eqnarray}
{\cal L}_{A}&=&-{1\over4}F_{\mu\nu}F^{\mu\nu}\:, 
\,\quad 
F_{\mu\nu}\equiv \partial_{\mu}A_{\nu}-\partial_{\nu}A_{\mu}\:, \\
{\cal L}_{\phi}&=& {1\over2}(\partial_{\mu}\phi-mA_{\mu})
(\partial^{\mu}\phi-mA^{\mu})\:, \\
{\cal L}_{\rm WZ}&=& -{k\over4}\epsilon^{\mu\nu\rho\sigma}
\phi F_{\mu\nu}F_{\rho\sigma}\:, \\ 
{\cal L}_{\psi}&=& \bar{\psi}i\gamma^{\mu} \!
\left[ \partial_{\mu}-ieA_{\mu} {1\over2}(1-\gamma_{5}) \right] \! \psi \:, 
\end{eqnarray} 
%(2.2)-(2.5)
%
where $A_{\mu}$ is a U(1) gauge field, $\phi$ a WZ scalar field, $\psi$ 
a Dirac field, and $m$, $k$ and $e$ are constants with suitable dimensions. 
[Our metric has signature $(+,\:-,\:-,\:-)$. The convention for the 
Levi--Civita symbol is $\epsilon^{0123}=-1$. The $\gamma_{5}$ matrix is  
defined by $\gamma_{5}(=\gamma_{5}^{\dagger})\equiv
i\gamma^{0}\gamma^{1}\gamma^{2}\gamma^{3}$.]   
The second term ${\cal L}_{\phi}$, which is the gauge-invariant mass term in 
the St\"uckelberg formalism, is necessary for the anomalous massive U(1) 
gauge theory to make a dynamical field of $\phi$ so that perturbative 
analysis can be applied to the theory. 
The lagrangian $\widetilde{\cal L}$ then describes a massive vector field  
interacting with chiral fermions. 
The WZ term ${\cal L}_{\rm WZ}$ is included in  
$\widetilde{\cal L}$ to cancel the chiral anomaly arising from  
the quantum  effects of $\psi\,$; 
although the lagrangian $\widetilde{\cal L}$ itself is not invariant under the 
gauge transformation   
\setcounter{enumi}{\value{equation}}
\addtocounter{enumi}{1}
\renewcommand{\theequation}{\thesection.\theenumi\alph{equation}}
\setcounter{equation}{0}
\begin{eqnarray}
\delta A_{\mu}&=& \partial_{\mu}\lambda \:, \\ 
\delta \phi&=& m\lambda \:, \\
\delta \psi&=& ie\lambda{1\over2}(1-\gamma_{5})\psi \:, \\
\delta \bar{\psi}&=& -ie\lambda\bar{\psi}{1\over2}(1+\gamma_{5}) \:,
\end{eqnarray}
%(2,6abcd)
%
the effective action 
\renewcommand{\theequation}{\thesection.\arabic{equation}}
\setcounter{equation}{\value{enumi}}
\begin{eqnarray}
\widetilde{\Gamma}[A_{\mu},\phi]=-i\ln \int {\cal D}\bar{\psi}{\cal D}\psi
\exp\!\left(i\int d^{4}x\widetilde{\cal L}\right)
\end{eqnarray}
%(2.7)
%
with $k=k_{0}\equiv e^{3}/(24\pi^{2}m)$ remains invariant 
due to the variation of the path-integral measure  
${\cal D}\bar{\psi}{\cal D}\psi$ [4].  
(Here the value of $k$ was chosen for the cancellation of 
the ^^ ^^ consistent" anomaly. 
If we discuss the cancellation of the ^^ ^^ covariant" anomaly, 
$k$ should be chosen to be $k=3k_{0}$.)
By virtue of the WZ term ${\cal L}_{\rm WZ}$, the U(1)  
gauge symmetry of the system holds at the quantum level. 
As a result, we can construct a consistent quantum field theory based 
on $\widetilde{\cal L}$, though this theory is power-counting  
nonrenormalizable owing to ${\cal L}_{\rm WZ}$.

To find a dual theory of the anomalous massive U(1) gauge theory, we now  
consider the first order lagrangian
\begin{eqnarray}
{\cal L}_{U}=-{1\over6}\epsilon^{\mu\nu\rho\sigma} U_{\mu}H_{\nu\rho\sigma}
+{1\over2}U_{\mu}U^{\mu}+{\cal L}_{BF} 
\end{eqnarray}
%(2.8)
%
with the so-called $BF$ term
\begin{eqnarray}
{\cal L}_{BF}={m\over4}\epsilon^{\mu\nu\rho\sigma}B_{\mu\nu}F_{\rho\sigma}
\:.
\end{eqnarray}
%(2.9)
%
Here $U_{\mu}$ is a vector field, $B_{\mu\nu}$ is an antisymmetric tensor  
field and 
\begin{eqnarray}
H_{\mu\nu\rho}\equiv F_{\mu\nu\rho}+k\omega_{\mu\nu\rho}\:,
\end{eqnarray}
%(2.10)
%
where
\begin{eqnarray}
F_{\mu\nu\rho} &\equiv& \partial_{\mu}B_{\nu\rho}
+\partial_{\nu}B_{\rho\mu}+\partial_{\rho}B_{\mu\nu} \:, \\
\omega_{\mu\nu\rho} &\equiv& A_{\mu}F_{\nu\rho}
+A_{\nu}F_{\rho\mu}+A_{\rho}F_{\mu\nu}\:.
\end{eqnarray}
%(2.11-12)
%
The tensor $\omega_{\mu\nu\rho}$ is nothing but the abelian 
Chern--Simons three-form. 
From ${\cal L}_{U}$ we obtain the Euler--Lagrange equation for 
$B_{\mu\nu\,}$: 
\begin{eqnarray}
\epsilon^{\mu\nu\rho\sigma}\partial_{\rho}(U_{\sigma}-mA_{\sigma})=0 \:,
\end{eqnarray}
%(2.13)
%
which can be solved as 
\begin{eqnarray}
U_{\mu}=mA_{\mu}-\partial_{\mu}\phi \:.
\end{eqnarray}
%(2.14)
%
Substituting (2.14) into (2.8) and removing total derivative terms, 
we arrive at ${\cal L}_{\phi}+{\cal L}_{\rm WZ}$. On the other hand, 
the Euler--Lagrange equation for $U_{\mu}$ is 
\begin{eqnarray}
U_{\mu}={1\over6}\epsilon_{\mu\nu\rho\sigma}H^{\nu\rho\sigma} \:.
\end{eqnarray}
%(2.15)
%
After substituting (2.15) into (2.8), we have 
${\cal L}_{H}+{\cal L}_{BF}$ with 
\begin{eqnarray}
{\cal L}_{H}={1\over12}H_{\mu\nu\rho}H^{\mu\nu\rho} \:.
\end{eqnarray}
%(2.16)
%
Therefore the lagrangian ${\cal L}_{H}+{\cal L}_{BF}$ is ^^ ^^ classically" 
equivalent to ${\cal L}_{\phi}+{\cal L}_{\rm WZ}$.

The field strength $H_{\mu\nu\rho}$ is invariant under the gauge 
transformation 
\setcounter{enumi}{\value{equation}}
\addtocounter{enumi}{1}
\renewcommand{\theequation}{\thesection.\theenumi\alph{equation}}
\setcounter{equation}{0}
\begin{eqnarray}
\delta A_{\mu} &=& \partial_{\mu}\lambda \:,  \\
\delta B_{\mu\nu} &=& 
\partial_{\mu}\xi_{\nu}-\partial_{\nu}\xi_{\mu}
-k\lambda F_{\mu\nu} \:.
\end{eqnarray}
%(2.17ab)
%
The gauge transformation rule (2.17b) and the field strength $H_{\mu\nu\rho}$ 
agree with the ones found in $N=1$ supergravity coupled to $N=1$  
supersymmetric Maxwell theory in ten dimensions [9]. 
Obviously ${\cal L}_{H}$ is gauge-invariant, whereas the topological term  
${\cal L}_{BF}$ is not gauge-invariant and transforms as
\renewcommand{\theequation}{\thesection.\arabic{equation}}
\setcounter{equation}{\value{enumi}}
\begin{eqnarray}
\delta{\cal L}_{BF}=-{1\over4}mk\epsilon^{\mu\nu\rho\sigma}
\lambda F_{\mu\nu}F_{\rho\sigma}
+\mbox{total derivative}\,.
\end{eqnarray}
%(2.18)
%
We notice that, up to the total derivative term, 
the transformation behavior of ${\cal L}_{BF}$ is exactly the same  
as that of ${\cal L}_{\rm WZ}$. The effective action 
\begin{eqnarray}
\Gamma[A_{\mu},B_{\mu\nu}]
=-i\ln \int {\cal D}\bar{\psi}{\cal D}\psi
\exp\!\left( i\int d^{4}x {\cal L}\right)
\end{eqnarray}
%(2.19)
%
with the lagrangian
\begin{eqnarray}
{\cal L}={\cal L}_{A}+{\cal L}_{H}+{\cal L}_{BF}+{\cal L}_{\psi}
\end{eqnarray}
%(2.20)
%
is thus gauge-invariant if $k=k_{0}$ for the consistent anomaly 
(or if $k=3k_{0}$ for the covariant anomaly). 
The lagrangian ${\cal L}$ defines a dual theory of the anomalous 
massive U(1) gauge theory described by $\widetilde{\cal L}\,$; 
these two theories  are equivalent (at least) at the ^^ ^^ classical" level. 
(As will be seen in the next section, this equivalence persists at the 
^^ ^^ quantum" level.)

The lagrangian ${\cal L}_{A}+{\cal L}_{H}+{\cal L}_{BF}$ with $k=0$ is known 
as a starting point of the topologically massive abelian gauge theory 
in four dimensions [7], in which the topological term ${\cal L}_{BF}$ makes  
$A_{\mu}$ massive.  
Here, we would like to emphasize that in the anomalous massive U(1) gauge 
theory defined by ${\cal L}$, the $BF$ term ${\cal L}_{BF}$  
plays the roles of the WZ term as well as the mass term of $A_{\mu}$.

\section{Duality at the quantum level}
\setcounter{equation}{0}
\vspace{0.5mm}
 
In order to complete our discussion in Section 2, 
we establish the equivalence of ${\cal L}_{H}+{\cal L}_{BF}$ and 
${\cal L}_{\phi}+{\cal L}_{\rm WZ}$ at the ^^ ^^ quantum" level. 
To this end, let us start with ${\cal L}_{H}+{\cal L}_{BF}$ and 
consider the covariant quantization of $B_{\mu\nu}$ using the BRST formalism 
[11]. We now introduce the following ghost and auxiliary 
fields associated with $B_{\mu\nu\,}$: anticommuting vector fields 
$\rho_{\mu}$ and $\bar{\rho}_{\mu}$, a commuting vector field $\beta_{\mu}$, 
anticommuting scalar fields $\chi$ and $\bar{\chi}$, and commuting scalar 
fields $\sigma$, $\varphi$ and $\bar{\sigma}$. For our 
discussion, we also need an anticommuting scalar ghost field $c$ associated 
with $A_{\mu}$. (In what follows $A_{\mu}$ is treated as an external classical 
field, and so it is not necessary to introduce further ghost and auxiliary  
fields.) The BRST transformation $\mbox{\boldmath{$\delta$}}$ is defined for 
$A_{\mu}$ and $B_{\mu\nu}$ by replacing the gauge parameters $\lambda$ and 
$\xi_{\mu}$ in (2.17) by the ghost fields $c$ and $\rho_{\mu\,}$:      
\setcounter{enumi}{\value{equation}}
\addtocounter{enumi}{1}
\renewcommand{\theequation}{\thesection.\theenumi\alph{equation}}
\setcounter{equation}{0}
\begin{eqnarray}
\mbox{\boldmath{$\delta$}} A_{\mu} &=& \partial_{\mu}c \:, 
\nonumber \\
\mbox{\boldmath{$\delta$}} B_{\mu\nu} &=& 
\partial_{\mu}\rho_{\nu}-\partial_{\nu}\rho_{\mu}
-kc F_{\mu\nu} \:.
\end{eqnarray}
%(3.1a) 
%
The BRST transformation rules of the other fields are defined so as to 
satisfy the nilpotency condition $\mbox{\boldmath{$\delta$}}^{2}=0\,$:
\begin{eqnarray}
\begin{array}{ll}
\vspace{1mm}
\mbox{\boldmath{$\delta$}}c=0 \:,  
&  \\
\vspace{1mm}
\mbox{\boldmath{$\delta$}}\rho_{\mu}=-i\partial_{\mu}\sigma \:, \quad
& \mbox{\boldmath{$\delta$}}\sigma=0 \:, \\
\vspace{1mm}
\mbox{\boldmath{$\delta$}}\bar{\rho}_{\mu}=i\beta_{\mu} \:, 
& \mbox{\boldmath{$\delta$}}\beta_{\mu}=0 \:, \\
\vspace{1mm}
\mbox{\boldmath{$\delta$}}\bar{\sigma}=\bar{\chi} \:, 
& \mbox{\boldmath{$\delta$}}\bar{\chi}=0 \:, \\
\vspace{1mm}
\mbox{\boldmath{$\delta$}}\varphi=\chi \:, 
& \mbox{\boldmath{$\delta$}}\chi=0 \:.
\end{array}
\end{eqnarray}
%(3.1b)
%
The ghost numbers assigned to 
$(A_{\mu},^{\,}c^{\,};\,B_{\mu\nu},^{\,}\rho_{\mu},^{\,}\bar{\rho}_{\mu},^{\,}
\sigma,^{\,}$ $\varphi,^{\,}\bar{\sigma}^{\,};^{\,}\beta_{\mu},^{\,}\chi,^{\,}
\bar{\chi})$ and 
$\mbox{\boldmath{$\delta$}}$ are  
$(0,^{\,}1;^{\,}0,^{\,}1,^{\,}$ $-1,^{\,}2,^{\,}0,^{\,}-2;^{\,}0,^{\,}$ 
$1,^{\,}-1)$ 
and $1$, respectively. 
To quantize $B_{\mu\nu}$, we have to introduce gauge-fixing terms with 
ghost number zero. Now we take the following gauge-fixing terms: 
\renewcommand{\theequation}{\thesection.\arabic{equation}}
\setcounter{equation}{\value{enumi}}
\begin{eqnarray}
{\cal L}_{\rm G1}=
-i\mbox{\boldmath{$\delta$}}[B_{\mu\nu}\partial^{\mu}\bar{\rho}^{\nu}] \:, 
\end{eqnarray}
%(3.2)
%
\begin{eqnarray}
{\cal L}_{\rm G2} &=&
i\mbox{\boldmath{$\delta$}}[\rho^{\mu}\partial_{\mu}\bar{\sigma}
+\bar{\rho}^{\mu}(\partial_{\mu}\varphi+uA_{\mu}+v\epsilon_{\mu\nu\rho\sigma}
\omega^{\nu\rho\sigma})] \:, 
\nonumber \\
& &
\end{eqnarray}
%(3.3)
%
where $u$ and $v$ are gauge parameters. 
Because of the nilpotency of $\mbox{\boldmath{$\delta$}}$, these gauge-fixing 
terms are invariant under the BRST transformation. 
The first term ${\cal L}_{\rm G1}$ breaks the gauge invariance 
of ${\cal L}_{H}$ explicitly.  
The second term ${\cal L}_{\rm G2}$ is necessary to break 
the invariance of ${\cal L}_{\rm G1}$ under the {\it secondary} gauge 
transformation $\delta\rho_{\mu}=\partial_{\mu}\varepsilon, \;
\delta\bar{\rho}_{\mu}=\partial_{\mu}\bar{\varepsilon}$. 
The gauge-fixing procedure for quantization of $B_{\mu\nu}$ is 
complete with ${\cal L}_{\rm G1}+{\cal L}_{\rm G2}$. 
Carrying out the BRST transformation in (3.2) and (3.3), we obtain 
\begin{eqnarray}
{\cal L}_{\rm G1}+{\cal L}_{\rm G2}&=&
-\beta^{\nu}(\partial^{\mu}B_{\mu\nu}+\partial_{\nu}\varphi
+uA_{\nu}+v\epsilon_{\nu\mu\rho\sigma}\omega^{\mu\rho\sigma})
\nonumber \\
& &
-i\bar{\rho}^{\nu}(\raisebox{-0.5ex}{\mbox{\Large$\Box$}}\rho_{\nu}
-\partial_{\nu}\partial^{\mu\!}\rho_{\mu}
-k\partial^{\mu}(cF_{\mu\nu})
\nonumber \\
& & 
+\partial_{\nu}\chi 
+u\partial_{\nu}c 
+3v\epsilon_{\nu\mu\rho\sigma}F^{\mu\rho}\partial^{\sigma}c)
\nonumber \\
& &
-i\rho^{\nu}\partial_{\nu}\bar{\chi}
-\bar{\sigma}\raisebox{-0.5ex}{\mbox{\Large$\Box$}}\sigma
+\mbox{total derivative}\,,
\nonumber \\
& &
\end{eqnarray} 
%(3.4)
%
where $\raisebox{-0.5ex}{\mbox{\Large$\Box$}}\equiv\partial_{\mu}
\partial^{\mu}$.

Let us now show the quantum equivalence of 
${\cal L}_{H}+{\cal L}_{BF}$ and ${\cal L}_{\phi}+{\cal L}_{\rm WZ}$.  
Consider the vacuum-to-vacuum amplitude 
\begin{eqnarray}
Z&=&\int{\cal D}{\cal M}\exp\biggl[ \, i\int d^{4}x
({\cal L}_{H}+{\cal L}_{BF}+{\cal L}_{\rm G1}+{\cal L}_{\rm G2}) \biggr] 
\nonumber \\
& &
\end{eqnarray}
%(3.5)
%
with the path-integral measure 
\begin{eqnarray}
{\cal D}{\cal M}
\equiv{\cal D}B_{\mu\nu}{\cal D}\rho_{\mu}{\cal D}\bar{\rho}_{\mu}
{\cal D}\beta_{\mu}{\cal D}\chi{\cal D}\bar{\chi}
{\cal D}\sigma{\cal D}\bar{\sigma}{\cal D}\varphi \:.
\end{eqnarray}
%(3.6)
%
We first notice that the integration over $\chi$ yields 
the delta-function $\prod_{x}\delta(\partial^{\nu}\bar{\rho}_{\nu})$. 
This function enables us to remove the two terms 
$i\bar{\rho}^{\nu}\partial_{\nu}\partial^{\mu\!}\rho_{\mu}$ and 
$-iu\bar{\rho}^{\nu}\partial_{\nu}c$ from the exponent in (3.5). 
After removing them, we express the delta-function  
$\prod_{x}\delta(\partial^{\nu}\bar{\rho}_{\nu})$ in the form of the integral 
over $\chi$ again. 
Then, the integration over $\rho_{\mu}$ and $\bar{\rho}_{\mu}$  
yields $(\det\!\raisebox{-0.5ex}{\mbox{\Large$\Box$}})^{4}$.  
After the integrations over 
$\chi$, $\bar{\chi}$, $\sigma$ and $\bar{\sigma}$, 
the amplitude $Z$ can be written as 
\begin{eqnarray}
Z &=& 
N_{1}(\det\!\raisebox{-0.5ex}{\mbox{\Large$\Box$}})^{3} \!
\int{\cal D}B_{\mu\nu} {\cal D}\beta_{\mu} {\cal D}\varphi 
\nonumber \\
& & \times \exp\biggl[ \, i \int d^{4}x \biggl\{ 
-{1\over4}B_{\mu\nu}\raisebox{-0.5ex}{\mbox{\Large$\Box$}}B^{\mu\nu}
-{1\over2}B^{\nu\rho}\partial_{\rho}\partial^{\mu}B_{\mu\nu} 
\nonumber \\
& & -{k\over2} B_{\mu\nu}\partial_{\rho}\omega^{\rho\mu\nu}
+{1\over12}k^{2}\omega_{\mu\nu\rho}\omega^{\mu\nu\rho}
+{m\over4}\epsilon^{\mu\nu\rho\sigma}B_{\mu\nu}F_{\rho\sigma}
\nonumber \\
& &
\biggl. \Bigl.
-\beta^{\nu}(\partial^{\mu}B_{\mu\nu}+\partial_{\nu}\varphi
+uA_{\nu}+v\epsilon_{\nu\mu\rho\sigma}\omega^{\mu\rho\sigma})
\biggr\} \biggr] \;.
\nonumber \\
& &
\end{eqnarray}
%(3.7)
%
Here and hereafter, $N_{i}$ $(i=1,2,3)$ denote normalization constants.  
Since the integration over $\beta_{\mu}$ yields 
the delta-function $\prod_{x,\nu}\delta(\partial^{\mu}B_{\mu\nu}+
\partial_{\nu}\varphi+uA_{\nu}+v\epsilon_{\nu\mu\rho\sigma}
\omega^{\mu\rho\sigma})\,$, 
we replace $\partial^{\mu}B_{\mu\nu}$ in the second term of the exponent 
by $-(\partial_{\nu}\varphi+uA_{\nu}+
v\epsilon_{\nu\mu\rho\sigma}\omega^{\mu\rho\sigma})\,$. 
Then, the integration over $B_{\mu\nu}$ leads to 
\begin{eqnarray}
Z &=& N_{2}\int{\cal D}\beta_{\mu}{\cal D}\varphi\exp
\biggl[ \,i\int d^{4}x \biggl\{ -{1\over2}\beta_{\mu}\beta^{\mu}
\nonumber \\
& & -\beta^{\mu}\!\left( \partial_{\mu}\varphi
+{u\over2}A_{\mu}+{v\over2}\epsilon_{\mu\nu\rho\sigma}\omega^{\nu\rho\sigma}
\right)
\nonumber \\
& & +{1\over2}(\partial^{\mu}\beta_{\mu})
\raisebox{-0.5ex}{\mbox{\Large$\Box$}}^{-1} \partial^{\nu}
(-\beta_{\nu}+uA_{\nu}+v\epsilon_{\nu\lambda\rho\sigma}
\omega^{\lambda\rho\sigma})
\nonumber \\
& & +{1\over4} \!\left(\, {1\over4}u^{2}-m^{2} \right)\! F_{\mu\nu}
\raisebox{-0.5ex}{\mbox{\Large$\Box$}}^{-1} F^{\mu\nu}
\nonumber \\
& & -\epsilon^{\mu\nu\rho\sigma}F_{\mu\nu}\!
\left(\, {3\over8}uv F_{\rho\sigma}\raisebox{-0.5ex}{\mbox{\Large$\Box$}}^{-1}
\partial^{\lambda}A_{\lambda} \right.
\nonumber \\
& & \left. +{1\over4}mk \raisebox{-0.5ex}{\mbox{\Large$\Box$}}^{-1}
\partial^{\lambda}\omega_{\lambda\rho\sigma} \right)
\nonumber \\
& & +{1\over4}k^{2} \partial^{\rho}\omega_{\rho\mu\nu}
\raisebox{-0.5ex}{\mbox{\Large$\Box$}}^{-1} 
\partial_{\sigma}\omega^{\sigma\mu\nu}
+{1\over12}(k^{2}+9v^{2})\omega_{\mu\nu\rho}\omega^{\mu\nu\rho}
\nonumber \\
& &
-{9\over32}v^{2}\epsilon^{\kappa\lambda\mu\nu}
F_{\kappa\lambda}F_{\mu\nu}\raisebox{-0.5ex}{\mbox{\Large$\Box$}}^{-1}
(\epsilon^{\pi\rho\sigma\tau}F_{\pi\rho}F_{\sigma\tau}) 
\biggr\} \biggr] \:.
\end{eqnarray}
%(3.8)
%
Since the integration over $\varphi$ yields the delta-function 
$\prod_{x}\delta(\partial^{\mu}\beta_{\mu})$, the terms proportional to 
$\partial^{\mu}\beta_{\mu}$ can be removed from the exponent in (3.8).  
Carrying out the integration over $\beta_{\mu}$, we obtain 
\begin{eqnarray}
Z &=& N_{3}\int{\cal D}\varphi
\exp\biggl[ \,i\int d^{4}x 
\nonumber \\
& & \times \biggl\{ {1\over2}\!
\left( \partial_{\mu}\varphi+{u\over2}A_{\mu}\right)\!\! 
\left( \partial^{\mu}\varphi+{u\over2}A^{\mu}\right) 
\nonumber \\
& & -{3\over4}v \epsilon^{\mu\nu\rho\sigma}
\varphi F_{\mu\nu}F_{\rho\sigma}
+{1\over4} \!\left( \, {1\over4}u^{2}-m^{2} \right)\! F_{\mu\nu}
\raisebox{-0.5ex}{\mbox{\Large$\Box$}}^{-1} F^{\mu\nu}
\nonumber \\
& & 
-{1\over8}(2mk+3uv)\partial^{\lambda}A_{\lambda}
\raisebox{-0.5ex}{\mbox{\Large$\Box$}}^{-1} 
(\epsilon^{\mu\nu\rho\sigma}F_{\mu\nu}F_{\rho\sigma})
\nonumber \\
& &
+{1\over32}(k^{2}-9v^{2})\epsilon^{\kappa\lambda\mu\nu}
F_{\kappa\lambda}F_{\mu\nu}\raisebox{-0.5ex}{\mbox{\Large$\Box$}}^{-1}
(\epsilon^{\pi\rho\sigma\tau}F_{\pi\rho}F_{\sigma\tau}) 
\biggr\} \biggr] \:, 
\nonumber \\
& &
\end{eqnarray}
%(3.9)
%
where we have used the following formulas: 
\begin{eqnarray}
\!\!\!& &\!\!\!
\partial^{\rho}\omega_{\rho\mu\nu}
\raisebox{-0.5ex}{\mbox{\Large$\Box$}}^{-1} 
\partial_{\sigma}\omega^{\sigma\mu\nu}
+{1\over3}\omega_{\mu\nu\rho}\omega^{\mu\nu\rho}
\nonumber \\  
\!\!\!& & 
\quad={1\over8}\epsilon^{\kappa\lambda\mu\nu}
F_{\kappa\lambda}F_{\mu\nu}\raisebox{-0.5ex}{\mbox{\Large$\Box$}}^{-1}
(\epsilon^{\pi\rho\sigma\tau}F_{\pi\rho}F_{\sigma\tau}) \quad
\nonumber \\
\!\!\!& & 
\,\qquad +\:\mbox{total derivative}\,,
\end{eqnarray}
%(3.10) 
%
\begin{eqnarray}
\epsilon^{\mu\nu\rho\sigma}\partial^{\lambda}\omega_{\lambda\mu\nu}
\raisebox{-0.5ex}{\mbox{\Large$\Box$}}^{-1}F_{\rho\sigma}
&=&
\partial^{\lambda}A_{\lambda}
\raisebox{-0.5ex}{\mbox{\Large$\Box$}}^{-1} 
(\epsilon^{\mu\nu\rho\sigma}F_{\mu\nu}F_{\rho\sigma}) 
\nonumber \\
& &
+\:\mbox{total derivative}\,.
\end{eqnarray}
%(3.11)
%
The first formula is due to the identity 
$\epsilon_{\kappa\lambda\mu\nu}\epsilon^{\pi\rho\sigma\tau}=
-{\delta_{\kappa}}^{[\pi}{\delta_{\lambda}}^{\rho}{\delta_{\mu}}^{\sigma}
{\delta_{\nu}}^{\tau]}$. 
The second formula is derived from 
$(\raisebox{-0.5ex}{\mbox{\Large$\Box$}}^{-1}A_{\lambda})
\epsilon^{[\lambda\mu\nu\rho}\partial^{\sigma]}\partial_{\mu}
\omega_{\nu\rho\sigma}=0\,$. (Here the brackets $[\;\:]$ indicate a total 
antisymmetrization with respect to all indices put between the brackets.)

The amplitude $Z$ is independent of the gauge parameters $u$ and $v$,  
as might be expected; the result of functional integration 
in (3.9) does not include these parameters. 
Taking into account the gauge independence of $Z$, we now choose $u$ 
and $v$ to be $u=-2m$ and $v=k/3$ so that all the non-local terms in (3.9)  
can vanish. 
Then (3.9) becomes 
\begin{eqnarray}
Z=N_{3}\int {\cal D}\phi \exp \biggl[  \,i\int d^{4}x 
({\cal L}_{\phi}+{\cal L}_{\rm WZ}) \biggr] \:,
\end{eqnarray}
%(3.12)
%
where $\varphi$ has been replaced by $\phi$. 
Therefore the vacuum-to-vacuum amplitude based on 
${\cal L}_{H}+{\cal L}_{BF}+{\cal L}_{\rm G1}+{\cal L}_{\rm G2}$ 
can be written as that based on ${\cal L}_{\phi}+{\cal L}_{\rm WZ}$, 
which demonstrates the equivalence of ${\cal L}_{H}+{\cal L}_{BF}$ and 
${\cal L}_{\phi}+{\cal L}_{\rm WZ}$ at the ^^ ^^ quantum" level. 
The two anomalous gauge theories characterized by ${\cal L}$ and 
$\widetilde{\cal L}$ are thus dual to each other not only at 
the classical level but also at the quantum level.  
Duality between the topologically massive abelian gauge theory 
and the abelian St\"uckelberg formalism is now obvious by setting $k=0$.

\section{Anomalous U(1) gauge theories in six dimensions} 
\setcounter{equation}{0}
\vspace{0.5mm}

We now consider the six-dimensional version of the lagrangian (2.20):
\begin{eqnarray}
{\cal L}_{1}^{(6)}={\cal L}_{A}^{(6)}+{\cal L}_{H1}^{(6)}
+{\cal L}_{BF1}^{(6)}+{\cal L}_{\psi}^{(6)} \:,
\end{eqnarray}
%(4.1)
%
where ${\cal L}_{A}^{(6)}$ and ${\cal L}_{\psi}^{(6)}$ are the six-dimensional 
analogs of ${\cal L}_{A}$ and ${\cal L}_{\psi\,}$, respectively. 
The remaining two terms are given by 
\begin{eqnarray}
{\cal L}_{H1}^{(6)} &=& {1\over2\!\cdot\!5!}
H_{\mu\nu\rho\sigma\tau} H^{\mu\nu\rho\sigma\tau} \:,
\\
& &
\nonumber \\
{\cal L}_{BF1}^{(6)} &=& {m\over48}
\epsilon^{\mu\nu\pi\rho\sigma\tau}B_{\mu\nu\pi\rho}F_{\sigma\tau}  
\end{eqnarray}
%(4.2)(4.3)
%
with a totally antisymmetric tensor field $B_{\mu\nu\pi\rho}$ and 
\begin{eqnarray}
H_{\mu\nu\rho\sigma\tau}\equiv F_{\mu\nu\rho\sigma\tau}
+\tilde{k}\omega_{\mu\nu\rho\sigma\tau} \:,
\end{eqnarray}
%(4.4)
%
where $\tilde{k}$ is a constant and 
\begin{eqnarray}
F_{\mu\nu\rho\sigma\tau} &\equiv& {1\over4!}
\partial_{\,[\mu}B_{\nu\rho\sigma\tau]} \:,
\nonumber \\
& &
\nonumber \\
\omega_{\mu\nu\rho\sigma\tau} &\equiv& {1\over8}
A_{[\mu}F_{\nu\rho}F_{\sigma\tau]} \:.
\end{eqnarray}
%(4.5)
%
[The conventions for the metric signature and the Levi--Civita symbol 
are $(+,\:-,\:-,\:-,\:-,\:-)$ and $\epsilon^{012345}=-1$.] 
The tensor $\omega_{\mu\nu\rho\sigma\tau}$ is known as the abelian 
Chern--Simons five-form. 
The lagrangian ${\cal L}_{1}^{(6)}$ describes an anomalous massive 
U(1) gauge theory in six dimensions. 
Since the field strength $H_{\mu\nu\rho\sigma\tau}$ 
is invariant under the gauge transformation 
\setcounter{enumi}{\value{equation}}
\addtocounter{enumi}{1}
\renewcommand{\theequation}{\thesection.\theenumi\alph{equation}}
\setcounter{equation}{0}
\begin{eqnarray}
\delta A_{\mu} &=& \partial_{\mu}\lambda \:, \\
\delta B_{\mu\nu\rho\sigma} &=& {1\over6}
\partial_{\,[\mu}\xi_{\nu\rho\sigma]}
-{1\over8}\tilde{k} \lambda F_{\,[\mu\nu}F_{\rho\sigma]} \:,
\end{eqnarray}
%(4.6ab)
%
${\cal L}_{H1}^{(6)}$ is also invariant, while the six-dimensional $BF$ term 
${\cal L}_{BF1}^{(6)}$ transforms as
\renewcommand{\theequation}{\thesection.\arabic{equation}}
\setcounter{equation}{\value{enumi}}
\begin{eqnarray}
\delta{\cal L}_{BF1}^{(6)} &=& -{1\over16}m \tilde{k} 
\epsilon^{\mu\nu\pi\rho\sigma\tau}
\lambda F_{\mu\nu}F_{\pi\rho}F_{\sigma\tau}
\nonumber \\
& & + \:\mbox{total derivative}\,.
\end{eqnarray}
%(4.7)
%
This transformation behavior is essential to cancellation of the 
six-dimensional chiral anomaly due to the quantum effects of 
the Dirac field $\psi\,$; 
the six-dimensional analog of (2.19) with the lagrangian  
${\cal L}^{(6)}_{1}$ is gauge-invariant, if $\tilde{k}$ 
is chosen to be $\tilde{k}=\tilde{k}_{0}\equiv e^{4}/(96\pi^{3}m)$ for 
the consistent anomaly (or $\tilde{k}=4\tilde{k}_{0}$ for the covariant 
anomaly).

Similarly to the case of four dimensions, we can represent 
${\cal L}_{H1}^{(6)}+{\cal L}_{BF1}^{(6)}$ as 
${\cal L}_{\phi}^{(6)}+{\cal L}_{\rm WZ}^{(6)}$ by means of a duality 
transformation at each of the classical and quantum levels, 
where ${\cal L}_{\phi}^{(6)}$ is the six-dimensional analog  
of (2.3) and ${\cal L}_{\rm WZ}^{(6)}$ is the WZ term in six 
dimensions: 
\begin{eqnarray}
{\cal L}_{\rm WZ}^{(6)}=-{{\tilde{k}}\over16}\epsilon^{\mu\nu\pi\rho\sigma\tau}
\phi F_{\mu\nu}F_{\pi\rho}F_{\sigma\tau} \:.
\end{eqnarray}
%(4.8)
%
The gauge theory defined by ${\cal L}_{1}^{(6)}$ 
is dual to the anomalous massive U(1) gauge theory defined by 
the lagrangian ${\cal L}_{A}^{(6)}+{\cal L}_{\phi}^{(6)}
+{\cal L}_{\rm WZ}^{(6)}+{\cal L}_{\psi}^{(6)}$.

In six dimensions, we can also consider an anomalous U(1) gauge 
theory with a topological term consisting of $F_{\mu\nu}$ and, 
instead of $B_{\mu\nu\rho\sigma}$, an antisymmetric tensor field $B_{\mu\nu}$ 
obeying the gauge transformation rule (2.17b). This theory is characterized 
by the lagrangian   
\begin{eqnarray}
{\cal L}_{2}^{(6)}={\cal L}_{A}^{(6)}+{\cal L}_{H2}^{(6)}
+{\cal L}_{BF2}^{(6)}+{\cal L}_{\psi}^{(6)} \:.
\end{eqnarray}
%(4.9)
%
Here ${\cal L}_{H2}^{(6)}$ has the same form as (2.16), and 
${\cal L}^{(6)}_{BF2}$ is the $BF^{2}$ term, a generalized $BF$ term, 
\begin{eqnarray}
{\cal L}_{BF2}^{(6)}={1\over16}l^{2}\epsilon^{\mu\nu\pi\rho\sigma\tau}
B_{\mu\nu}F_{\pi\rho}F_{\sigma\tau} \:,
\end{eqnarray}
%(4.10)
%
where $l$ is a constant. Under the gauge transformation (2.17), 
$\,{\cal L}_{BF2}^{(6)}$ transforms in a similar manner to 
${\cal L}_{BF1}^{(6)}$, namely 
\begin{eqnarray}
\delta{\cal L}_{BF2}^{(6)} &=& -{1\over16}l^{2} k 
\epsilon^{\mu\nu\pi\rho\sigma\tau}
\lambda F_{\mu\nu}F_{\pi\rho}F_{\sigma\tau}
\nonumber \\
& & + \:\mbox{total derivative}\,.
\end{eqnarray}
%(4.11)
%
Hence, the six-dimensional analog of (2.19) with the lagrangian 
${\cal L}_{2}^{(6)}$ is also gauge-invariant, if $k$ is chosen 
to be $k=\bar{k}_{0}\equiv e^{4}/(96\pi^{3}l^{2})$ for the consistent 
anomaly (or $k=4\bar{k}_{0}$ for the covariant anomaly).

The two anomalous gauge theories defined by ${\cal L}_{1}^{(6)}$ and 
${\cal L}_{2}^{(6)}$ are substantially different. 
The lagrangian ${\cal L}_{1}^{(6)}$ describes a massive vector field  
interacting with chiral fermions; the only physical degree of freedom of 
$B_{\mu\nu\rho\sigma}$ is observed as 
the longitudinal mode of the massive vector field. 
The $BF$ term ${\cal L}_{BF1}^{(6)}$ functions both as the WZ term and as 
the mass term of $A_{\mu}$. 
In contrast, the lagrangian ${\cal L}_{2}^{(6)}$ describes a massless 
vector field interacting with chiral fermions and with a massless rank-two 
antisymmetric tensor field.  All the physical degrees of 
freedom of $B_{\mu\nu}$ are observed as its own massless modes. 
Although ${\cal L}_{BF2}^{(6)}$ plays the role of the WZ term,   
it does not function as a mass term.

It has been shown that in the two theories defined by ${\cal L}_{1}^{(6)}$ and 
${\cal L}_{2}^{(6)}$, the six-dimensional chiral anomaly vanishes  
by virtue of the topological terms  
${\cal L}_{BF1}^{(6)}$ and ${\cal L}_{BF2}^{(6)}$ with 
suitable coefficients. We can thus construct consistent quantum  
field theories based on ${\cal L}_{1}^{(6)}$ and ${\cal L}_{2}^{(6)}$.

In higher dimensions, the varieties of generalized $BF$ terms increase: 
for example, in $2n$ dimensions, we can construct a $BF$ term and $n-2$ 
generalized $BF$ terms, 
\begin{eqnarray}
& &
\epsilon^{\mu_{1}\cdots\mu_{2p}\mu_{2p+1}\mu_{2p+2}\cdots\mu_{2n-1}\mu_{2n}}
\nonumber \\
& & \;\times
B_{\mu_{1}\cdots\mu_{2p}}F_{\mu_{2p+1}\mu_{2p+2}}\cdots F_{\mu_{2n-1}\mu_{2n}}
\:,   \quad \qquad 
\nonumber \\
%& & \nonumber \\
& &  \quad \qquad \qquad (p=1,\,2,\,3,\ldots,n-1)
\end{eqnarray}
%(4.12)
%
where $B_{\mu_{1}\cdots\mu_{2p}}$ is a totally antisymmetric tensor field of 
rank $2p$. Imposing an appropriate gauge transformation rule such as (2.17b) 
and (4.6b) on $B_{\mu_{1}\cdots\mu_{2p}\,}$, 
we can make the $BF$ and generalized $BF$ terms (4.12) 
function as the WZ term for the chiral U(1) anomaly in $2n$ dimensions. 

\section{Anomalous non-abelian gauge theory} 
\setcounter{equation}{0}
\vspace{0.5mm}

We next discuss an anomalous non-abelian gauge theory in four 
dimensions whose gauge group is SU(2)$\times$U(1). 
Introducing a U(1) gauge field $A_{\mu\,}$, a SU(2) gauge field 
$\widehat{A}_{\mu} 
={1\over2}{\widehat{A}_{\mu}}\raisebox{0.98ex}{\mbox{\scriptsize{$a$}}}
\sigma_{a}$ 
[$\,\sigma_{a}$ $(a=1,\,2,\,3)$ denote the Pauli matrices\,],  
and a Dirac field $\widehat{\psi}$ belonging to the fundamental  
representation of SU(2), we consider the lagrangian 
$\widehat{\cal L}_{A}+\widehat{\cal L}_{\psi}$ with   
\begin{eqnarray}
\widehat{\cal L}_{A} &=& -{1\over4}F_{\mu\nu}F^{\mu\nu}
-{1\over4}{\rm tr}[\widehat{F}_{\mu\nu}\widehat{F}^{\mu\nu}] \:, 
\\
\widehat{\cal L}_{\psi} &=& \bar{\widehat{\psi}}i\gamma^{\mu} \!
\left[ \partial_{\mu}-i(eA_{\mu}\sigma_{0}
+g\widehat{A}_{\mu}){1\over2}(1-\gamma_{5}) \right] \! 
\widehat{\psi} \:, 
\end{eqnarray} 
%(5.1)(5.2)
%
where 
\begin{eqnarray}
F_{\mu\nu} &\equiv& 
\partial_{\mu}A_{\nu}-\partial_{\nu}A_{\mu} \:, \\ 
\widehat{F}_{\mu\nu} &\equiv& \partial_{\mu}\widehat{A}_{\nu}
-\partial_{\nu}\widehat{A}_{\mu}-ig[\widehat{A}_{\mu},\,\widehat{A}_{\nu}]\:, 
\end{eqnarray}
%(5.3)(5.4)
%
$\sigma_{0}$ is the $2\times2$ unit matrix, and $e$ and $g$ are constants.   
The lagrangian $\widehat{\cal L}_{A}+\widehat{\cal L}_{\psi}$ is invariant 
under the gauge transformation 
\setcounter{enumi}{\value{equation}}
\addtocounter{enumi}{1}
\renewcommand{\theequation}{\thesection.\theenumi\alph{equation}}
\setcounter{equation}{0}
\begin{eqnarray}
\delta A_{\mu}&=&\partial_{\mu}\lambda \:, \\ 
\delta \widehat{A}_{\mu}&=&\partial_{\mu}\widehat{\lambda}
-ig[\widehat{A}_{\mu},\,\widehat{\lambda}^{\,}] \:, \\ 
\delta \widehat{\psi}&=&
i(e\lambda\sigma_{0}+g\widehat{\lambda})
{1\over2}(1-\gamma_{5})\widehat{\psi} \:, \\
\delta \bar{\widehat{\psi}}&=& -i\bar{\widehat{\psi}}
(e\lambda\sigma_{0}+g\widehat{\lambda})
{1\over2}(1+\gamma_{5}) \:,
\end{eqnarray}
%(5,5abcd)
%
where $\widehat{\lambda}$ is represented as $\widehat{\lambda}
={1\over2}\widehat{\lambda}\raisebox{0.94ex}{\mbox{\scriptsize{$a$}}}
\sigma_{a\,}$.

In the chiral gauge theory defined by  
$\widehat{\cal L}_{A}+\widehat{\cal L}_{\psi\,}$, a non-abelian chiral 
anomaly arises necessarily due to the quantum effects of $\widehat{\psi}\,$. 
We can find the anomaly in the gauge transformation of the 
effective action $W[A_{\mu}, \widehat{A}_{\mu}]$ defined from the 
path-integral of $\exp(i\int d^{4}x \widehat{\cal L}_{\psi})$ 
over $\bar{\widehat{\psi}}$ and $\widehat{\psi}\,$.  
The gauge transformation of $W$ is systematically calculated using 
the perturbative or non-perturbative method with suitable 
regularization procedures for ill-defined quantities [12]. 
Adopting a certain regularization procedure,  
we obtain the ^^ ^^ consistent" anomaly; 
in the case at hand, it can be written as 
\renewcommand{\theequation}{\thesection.\arabic{equation}}
\setcounter{equation}{\value{enumi}}
\begin{eqnarray}
\delta W &=& 
\int d^{4}x {1\over{24\pi^{2}}} \epsilon^{\mu\nu\rho\sigma}
\nonumber \\
\!\!\!& &\times{\rm tr} \biggl[ \Lambda\partial_{\mu}\!
\left( {\cal A}_{\nu}\partial_{\rho}{\cal A}_{\sigma}
-{i\over2}{\cal A}_{\nu}{\cal A}_{\rho}{\cal A}_{\sigma} \right) \biggr]
\nonumber \\
&=&
\int d^{4}x {1\over{24\pi^{2}}}  
\epsilon^{\mu\nu\rho\sigma}
\biggl\{ {1\over2}e^{3}\lambda F_{\mu\nu}F_{\rho\sigma}
\nonumber \\
& &
+eg^{2}F_{\mu\nu} 
{\rm tr}[\widehat{\lambda}\partial_{\rho}\widehat{A}_{\sigma}]
-{i\over2}eg^{3}{\rm tr}[\widehat{\lambda} \partial_{\mu}
(A_{\nu}\widehat{A}_{\rho}\widehat{A}_{\sigma})]
\nonumber \\
& &
+eg^{2}\lambda{\rm tr} \biggl[ \partial_{\mu}\!
\left( \widehat{A}_{\nu}\partial_{\rho}
\widehat{A}_{\sigma}-{i\over2}g\widehat{A}_{\nu}\widehat{A}_{\rho}
\widehat{A}_{\sigma} \right) \biggr]  
\nonumber \\
& &
+g^{3}{\rm tr} \biggl[ \widehat{\lambda}\partial_{\mu} \!\left( 
\widehat{A}_{\nu}\partial_{\rho}
\widehat{A}_{\sigma}-{i\over2}g\widehat{A}_{\nu}\widehat{A}_{\rho}
\widehat{A}_{\sigma} \right) \biggr] \biggr\} \:,
\end{eqnarray}   
%(5.6)
% 
where ${\cal A}_{\mu}\equiv eA_{\mu}\sigma_{0}+g\widehat{A}_{\mu}$ 
and $\Lambda\equiv e\lambda\sigma_{0}+g\widehat{\lambda}\,$. 
The term in the last line of (5.6) vanishes because of a property 
of the Pauli matrices. The remainder of the integrand can be expressed as  
\begin{eqnarray}
& &{1\over{24\pi^{2}}} \epsilon^{\mu\nu\rho\sigma}
\biggl\{ {1\over2}e^{3}\lambda F_{\mu\nu}F_{\rho\sigma}
+{3\over2}eg^{2}F_{\mu\nu} 
{\rm tr}[\widehat{\lambda}\partial_{\rho}\widehat{A}_{\sigma}] \biggr\}
\nonumber \\ 
& &-\delta w +\mbox{total derivative}
\end{eqnarray}
%(5.7)
%
with
\begin{eqnarray}
w \equiv
{1\over{24\pi^{2}}}eg^{2}\epsilon^{\mu\nu\rho\sigma\!}
A_{\mu}{\rm tr}\biggl[ \widehat{A}_{\nu}
\partial_{\rho}\widehat{A}_{\sigma}
-{i\over2}g\widehat{A}_{\nu}\widehat{A}_{\rho}
\widehat{A}_{\sigma} \biggr] \:.  
\end{eqnarray}
%(5.8)
%
In (5.7), $-\delta w$ is considered to be a ^^ ^^ trivial" 
violation of gauge symmetry, 
since it can be removed by adding the ^^ ^^ local" functional 
$\int d^{4}x w$ to the effective action $W$. 
(The trivial violation $-\delta w$ is due to the regularization procedure  
adopted here, and so does not alter the anomalous content of the theory.) 
We thus arrive at the anomaly written in the following form: 
\begin{eqnarray}
& & \delta\!\left(W+\int d^{4}x w\right)
\nonumber \\
& & \:\;\; =\int d^{4}x {1\over{24\pi^{2}}}
\epsilon^{\mu\nu\rho\sigma}
\biggl\{ {1\over2}e^{3}\lambda F_{\mu\nu}F_{\rho\sigma}
\biggr. \nonumber \\ 
& & \qquad \biggl.
+{3\over2}eg^{2}F_{\mu\nu} 
{\rm tr}[\widehat{\lambda}\partial_{\rho}\widehat{A}_{\sigma}] \biggr\} \:.
\end{eqnarray}
%(5.9)
%
The right-hand side of (5.9) can never be written in the form of the gauge 
variation of a local functional only in $A_{\mu}$ and $\widehat{A}_{\mu\,}$.  
For this reason, 
we can not find a gauge-invariant effective action without introducing 
extra physical degrees of freedom.

In order that the gauge symmetry may be restored to the theory, 
let us introduce the $BF$ term 
(2.9) with $B_{\mu\nu}$ obeying the gauge transformation rule
\begin{eqnarray}
\delta B_{\mu\nu}=\partial_{\mu}\xi_{\nu}-\partial_{\nu}\xi_{\mu}
-k\lambda F_{\mu\nu}-\widehat{k}{\rm tr}[\widehat{\lambda}
\partial_{\,[\mu}\widehat{A}_{\nu]}] 
\end{eqnarray}
%(5.10)
%
instead of (2.17b). Here $k$ and $\widehat{k}$ are constants. 
At the same time, we modify the field strength 
$H_{\mu\nu\rho}$ in (2.10) so it is invariant under the gauge 
transformation defined by (5.5) and (5.10). 
The field strength modified satisfactorily is found to be 
\begin{eqnarray}
\widehat{H}_{\mu\nu\rho}\equiv F_{\mu\nu\rho}+k\omega_{\mu\nu\rho}
+\widehat{k}\widehat{\omega}_{\mu\nu\rho} \:,
\end{eqnarray}
%(5.11)
%
where $F_{\mu\nu\rho}$ and $\omega_{\mu\nu\rho}$ have been given in (2.11) 
and (2.12), and $\widehat{\omega}_{\mu\nu\rho}$ is the non-abelian 
Chern--Simons three-form   
\begin{eqnarray}
\widehat{\omega}_{\mu\nu\rho}\equiv {\rm tr}
\biggl[ \widehat{A}_{[\mu}\partial_{\nu}\widehat{A}_{\rho]}
-{2\over3}ig\widehat{A}_{[\mu}\widehat{A}_{\nu}\widehat{A}_{\rho]} \biggr]\:.
\end{eqnarray}
%(5.12)
%
The gauge transformation rule (5.10) and the field strength 
$\widehat{H}_{\mu\nu\rho}$ were first found by Chapline and Manton in 
the study of $N=1$ supergravity coupled to $N=1$ supersymmetric Yang--Mills 
theory in ten dimensions [13]. 
Recently, (5.10) and $\widehat{H}_{\mu\nu\rho}$ 
have also been obtained in the Yang--Mills theory in loop space 
with the affine gauge group [14]. 
Since $\widehat{H}_{\mu\nu\rho}$ is gauge-invariant, 
\begin{eqnarray}
\widehat{\cal L}_{H}={1\over12}\widehat{H}_{\mu\nu\rho}
\widehat{H}^{\mu\nu\rho} 
\end{eqnarray}
%(5.13)
%
is also invariant, while the $BF$ term (2.9) 
transforms as  
\begin{eqnarray}
\delta{\cal L}_{BF}&=& 
-{1\over4}m\epsilon^{\mu\nu\rho\sigma}
(k\lambda F_{\mu\nu}F_{\rho\sigma}
+2\widehat{k}F_{\mu\nu} 
{\rm tr}[\widehat{\lambda}\partial_{\rho}\widehat{A}_{\sigma}]) 
\nonumber \\
& & +\:\mbox{total derivative}\,.
\end{eqnarray}
%(5.14)
%
Comparing (5.14) with (5.9), we see that the non-abelian analog of (2.19) 
with the lagrangian 
\begin{eqnarray}
\widehat{\cal L}=\widehat{\cal L}_{A}+\widehat{\cal L}_{H}
+{\cal L}_{BF}+\widehat{\cal L}_{\psi} 
\end{eqnarray}
%(5.15)
%
is gauge-invariant (up to the trivial violation $-\delta w$), 
if $k$ and $\widehat{k}$ are chosen to be 
$k=e^{3}/(12\pi^{2}m)$ and $\widehat{k}=eg^{2}/(8\pi^{2}m)$.  
In this case, the $BF$ term ${\cal L}_{BF}$ plays the roles of the WZ term 
for the non-abelian chiral anomaly as well as the mass term of $A_{\mu}$. 
Consequently, the SU(2)$\times$U(1) gauge symmetry is restored 
at the quantum level, though the U(1) gauge field $A_{\mu}$ becomes massive.   
(The SU(2) gauge field $\widehat{A}_{\mu}$ remains massless.)
We can thus construct a consistent quantum theory based on $\widehat{\cal L}$.

Through the first order lagrangian that is defined by replacing  
$H_{\mu\nu\rho}$ in (2.8) with $\widehat{H}_{\mu\nu\rho}$, we can easily 
verify the classical equivalence of 
$\widehat{\cal L}_{H}+{\cal L}_{BF}$ and 
\begin{eqnarray}
{\cal L}'={\cal L}_{\phi}+\widehat{\cal L}_{\rm WZ}
-{1\over6}m\widehat{k}\epsilon^{\mu\nu\rho\sigma\!}A_{\mu}
\widehat{\omega}_{\nu\rho\sigma} \:,
\end{eqnarray}
%(5.16)
%
where ${\cal L}_{\phi}$ is given by (2.3) and $\widehat{\cal L}_{\rm WZ}$ 
is the WZ term
\begin{eqnarray}
\widehat{\cal L}_{\rm WZ}=-{1\over4}\epsilon^{\mu\nu\rho\sigma}
\phi(kF_{\mu\nu}F_{\rho\sigma}
+\widehat{k}{\rm tr}[\widehat{F}_{\mu\nu}\widehat{F}_{\rho\sigma}]) \:.
\end{eqnarray}
%(5.17)
%
The equivalence of $\widehat{\cal L}_{H}+{\cal L}_{BF}$ and ${\cal L}'$ 
holds also at the quantum level.  
To show this, it is necessary to introduce an anticommuting scalar ghost field 
$\widehat{c}={1\over2}\widehat{c}\raisebox{0.85ex}{\mbox{\scriptsize{$\,a$}}}
\sigma_{a}$ associated with 
$\widehat{A}_{\mu}$ in addition to the ghost and auxiliary fields 
introduced in Section 3. From the gauge transformation rules (5.5a), (5.5b) 
and (5.10), the BRST transformation rules of  
$A_{\mu}$, $\widehat{A}_{\mu}$ and $B_{\mu\nu}$ are determined to be 
\setcounter{enumi}{\value{equation}}
\addtocounter{enumi}{1}
\renewcommand{\theequation}{\thesection.\theenumi\alph{equation}}
\setcounter{equation}{0}
\begin{eqnarray}
\mbox{\boldmath{$\delta$}} A_{\mu} &=& \partial_{\mu}c \:, 
\nonumber \\ 
\mbox{\boldmath{$\delta$}} \widehat{A}_{\mu} &=& 
\partial_{\mu}\widehat{c}-ig[\widehat{A}_{\mu},\,\widehat{c}^{\,}] \:, 
\nonumber \\ 
\mbox{\boldmath{$\delta$}} B_{\mu\nu}
&=& \partial_{\mu}\rho_{\nu}-\partial_{\nu}\rho_{\mu}
-kcF_{\mu\nu}-\widehat{k}{\rm tr}[\widehat{c}_{\,}
\partial_{\,[\mu}\widehat{A}_{\nu]}] \:.
\end{eqnarray}
%(5.18a)
%
Here the ghost number 1 has been assigned to $\widehat{c\,}$. 
The BRST transformation rule of $\widehat{c}$ is determined from 
the nilpotency of $\mbox{\boldmath{$\delta$}}$, which also demands some 
modifications of the BRST transformation rules of $\rho_{\mu}$ and 
$\sigma$ given in (3.1b) [15]. We thus have
\begin{eqnarray}
\mbox{\boldmath{$\delta$}} \widehat{c} &=&
ig\widehat{c}
\raisebox{0.85ex}{\mbox{\scriptsize{$\,2$}}} \:,
\nonumber \\
\mbox{\boldmath{$\delta$}} \rho_{\mu} &=&
-i\partial_{\mu}\sigma -ig\widehat{k} {\rm tr}
[\widehat{c}\raisebox{0.85ex}{\mbox{\scriptsize{$\,2$}}}
\widehat{A}_{\mu}] \:,
\nonumber \\
\mbox{\boldmath{$\delta$}} \sigma &=&
-{1\over3}g\widehat{k} {\rm tr}
[\widehat{c}\raisebox{0.85ex}{\mbox{\scriptsize{$\,3$}}}] \:.
\end{eqnarray}
%(5.18b)
%
The other ghost and auxiliary fields obey the transformation rules in (3.1b). 
Now we take the gauge-fixing terms (3.2) and, instead of (3.3), 
\renewcommand{\theequation}{\thesection.\arabic{equation}}
\setcounter{equation}{\value{enumi}}
\begin{eqnarray}
\widehat{\cal L}_{\rm G2}&=&
i\mbox{\boldmath{$\delta$}}[\rho^{\mu}\partial_{\mu}\bar{\sigma}
+\bar{\rho}^{\mu}(\partial_{\mu}\varphi+uA_{\mu}
\nonumber \\
& &
+v\epsilon_{\mu\nu\rho\sigma}\omega^{\nu\rho\sigma}
+\widehat{v}\epsilon_{\mu\nu\rho\sigma}
\widehat{\omega}^{\nu\rho\sigma})] 
\end{eqnarray}
%(5.19)
%
with the gauge parameters $u$, $v$ and $\widehat{v}$. 
Starting with the vacuum-to-vacuum amplitude, $\widehat{Z}$, with the 
lagrangian $\widehat{\cal L}_{H}+{\cal L}_{BF}+{\cal L}_{\rm G1}
+\widehat{\cal L}_{\rm G2}$ and by following the same procedure that 
we used in Section 3, we can rewrite $\widehat{Z}$ as the vacuum-to-vacuum 
amplitude with a lagrangian consisting of certain local and non-local terms. 
Since the amplitude $\widehat{Z}$ is independent of the gauge parameters,  
we choose them to be $u=-2m$, $v=k/3$ and $\widehat{v}=\widehat{k}/3\,$; 
then all the non-local terms vanish and ${\cal L}'$ 
alone remains as a lagrangian defining $\widehat{Z}$. 
This result demonstrates the quantum equivalence of 
$\widehat{\cal L}_{H}+{\cal L}_{BF}$ and ${\cal L}'$. 
Needless to say, the lagrangian $\widehat{\cal L}_{A}+{\cal L}'
+\widehat{\cal L}_{\psi}$ defines a dual theory of the anomalous 
gauge theory described by $\widehat{\cal L}$.

Adding a gauge variation of any local functional in gauge fields to 
an anomaly does not alter the anomalous content of the theory. 
In other words, an anomaly is unique up to gauge variations of 
local functionals in gauge fields, and so may take various forms. 
In the theory under consideration, adding the local functional 
$\int d^{4}x(w-{1\over6}m\widehat{k}\epsilon^{\mu\nu\rho\sigma\!}A_{\mu}
\widehat{\omega}_{\nu\rho\sigma})$ to the effective action $W$, 
we obtain a form of the anomaly that is completely canceled with  
the gauge variation of the WZ action 
$\int d^{4}x\widehat{\cal L}_{\rm WZ}$. 
Hence, it is possible to construct a consistent quantum theory 
based on the lagrangian 
$\widehat{\cal L}_{A}+{\cal L}_{\phi}+\widehat{\cal L}_{\rm WZ}
+\widehat{\cal L}_{\psi}$.

Without any essential modification, our discussion in this section  
is applicable to the anomalous gauge theory 
with the gauge group $G$$\times$U(1) whenever the generators, $T_{a}$, of 
$G$ satisfy the conditions 
${\rm tr}[T_{a}]=0$ and ${\rm tr}[T_{a}\{T_{b},\,T_{c}\}]=0$ [16]. 
For example, the generators of SO($n$) $(n\geq3,\:n\neq6)$ satisfy 
these conditions.

\section{Summary and Discussion}

In this paper we have studied anomalous gauge theories 
in four and six dimensions that contain antisymmetric tensor fields. 
It has been shown that in the anomalous U(1) and SU(2)$\times$U(1)  
gauge theories in four dimensions, the $BF$ term with an antisymmetric  
tensor field $B_{\mu\nu}$ plays the roles of 
the WZ terms as well as the mass term of the U(1) gauge field $A_{\mu}$  
by imposing suitable gauge transformation rules on $B_{\mu\nu}$. 
By virtue of the $BF$ term, the chiral anomalies are 
canceled and the gauge symmetries are recovered to the theories.

We have demonstrated, both at the classical and quantum levels, that 
the four-dimensional anomalous U(1) gauge theory with $BF$ term is dual to  
the four-dimensional anomalous U(1) gauge theory with a WZ term 
and with a St\"uckelberg type mass term of $A_{\mu}$. 
Similar duality has also been discussed 
in the six-dimensional anomalous U(1) gauge theory with $BF$ term and 
in the four-dimensional anomalous SU(2)$\times$U(1) gauge theory with $BF$ 
term.

In six dimensions, we have considered the anomalous U(1) gauge theory 
with $BF^{2}$ term. 
This theory is substantially different from the six-dimensional 
anomalous U(1) gauge theory with $BF$ term, 
since the $BF$ term functions both as the WZ term and as the mass term 
of $A_{\mu}$, while the $BF^{2}$ term functions only as the WZ term. 
In each theory, the $BF$ or $BF^{2}$ term restores the U(1) gauge symmetry 
at the quantum level.

The $BF$ terms are known as generalizations 
of the Chern--Simons term in three dimensions;  
the anomalous gauge theories with $BF$ terms might be formulated to be  
higher-dimensional analogs of the three-dimensional anomalous gauge theory  
in which the Chern--Simons term 
restores the gauge symmetry at the quantum level [17].

Power-counting renormalizability of the four-dimensional anomalous 
U(1) gauge theory with WZ term is spoiled because of the WZ term, since 
it has a proportional constant with dimensions of length.   
On the other hand, the $BF$ term has a proportional constant with dimensions 
of mass, although it functions as the WZ term. This is desirable to 
renormalizability of the four-dimensional anomalous U(1) gauge theory with 
$BF$ term. However, instead of the $BF$ term, 
the Chern--Simons three-form included in the field strength 
$H_{\mu\nu\rho}$ has a proportional constant with dimensions of length. 
Consequently, the four-dimensional anomalous U(1) gauge theory with $BF$ term 
is also power-counting nonrenormalizable. 
The same can be said of the four-dimensional anomalous SU(2)$\times$U(1)  
gauge theory with $BF$ term.

The anomalous SU(2)$\times$U(1) gauge theory with $BF$ term seems to be 
applicable to constructing an electroweak model in which anomalies 
due to quarks and leptons do not cancel among these particles. 
An anomalous SU(2)$\times$U(1) gauge theory with WZ term has indeed be applied 
to the description of electroweak model that lacks the top quark [18]. 
In our theory, however, only the U(1) gauge field $A_{\mu}$ becomes massive, 
while the SU(2) gauge field $\widehat{A}_{\mu}$ remains massless. 
This is due to the fact that the only abelian antisymmetric tensor field 
$B_{\mu\nu}$ has been introduced into the theory. 
The gauge fields $A_{\mu}$ and $\widehat{A}_{\mu}$ (or their linear  
combinations) can not be identified with the gauge bosons in 
the Weinberg--Salam  model without considering further physical degrees of 
freedom. If we describe those gauge bosons in terms of the anomalous  
SU(2)$\times$U(1) gauge theory with $BF$ term, it will be necessary to 
introduce non-abelian antisymmetric tensor fields of rank two [8] and  
to consider their interactions with the gauge fields $A_{\mu}$ and  
$\widehat{A}_{\mu}$. 
\vspace{4mm}

\noindent
{\bf Acknowledgements}

\vspace{2mm}

We are grateful to Professor S. Naka and other members of the Theoretical  
Physics Group at Nihon University for their encouragements and useful 
comments. This work was supported in part by Nihon University Research Grant.

\vspace{1cm}
%\newpage

\begin{center}
{\Large\bf References}

\vspace{1mm}

\end{center}
\begin{enumerate}

%[1]
\item R. Jackiw and R. Rajaraman, Phys. Rev. Lett. {\bf 54}, 1219 (1985); 
{\bf 54}, 2060(E) (1985). 
%
%[2]
\item N. V. Krasnikov, Nuovo Cimento {\bf 89A}, 308 (1985); {\bf 95A}, 325 
(1986). 

\vspace{1mm}

I. G. Halliday, E. Rabinovici, A. Schwimmer and M. Chanowitz, Nucl. Phys. 
{\bf B268}, 413 (1986). 

\vspace{1mm}

M. S. Chanowitz, Phys. Lett. {\bf B171}, 280 (1986). 

\vspace{1mm}

S. Miyake and K. Shizuya, Phys. Rev. {\bf D36}, 3781 (1987); 
Phys. Rev. {\bf D37}, 2282 (1988). 
%
%[3]
\item O. Babelon, F. A. Schaposnik and C. M. Viallet, Phys. Lett. {\bf B177}, 
385 (1986). 

\vspace{1mm}

K. Harada and I. Tsutsui, Phys. Lett. {\bf B183}, 311 (1987); Prog. Theor. 
Phys. {\bf 78}, 878 (1987). 
%
%[4]
\item L. D. Faddeev and S. L. Shatashvili, Phys. Lett. {\bf B167}, 225 (1986). 
%
%[5]
\item R. Rajaraman, Phys. Lett. {\bf B184}, 369 (1987). 

\vspace{1mm}

A. Andrianov, A. Bassetto and R. Soldati, Phys. Rev. Lett. {\bf 63}, 1554 
(1989). 

\vspace{1mm}

S. Miyake and K. Shizuya, Mod. Phys. Lett. {\bf A4}, 2675 (1989).  
%
%[6]
\item A. Della Selva, L. Masperi and G. Thompson, Phys. Rev. {\bf D37}, 
2347 (1988). 

\vspace{1mm}

T. Fujiwara, Y. Igarashi and J. Kubo, Nucl. Phys. {\bf B341}, 695 (1990).   

%[7]
\item E. Cremmer and J. Scherk, Nucl. Phys. {\bf B72}, 117 (1974). 

\vspace{1mm}

A. Aurilia and Y. Takahashi, Prog. Theor. Phys. {\bf 66}, 693 (1981). 

\vspace{1mm}

I. Oda and S. Yahikozawa, Prog. Theor. Phys. {\bf 83}, 991 (1990). 

\vspace{1mm}

T. J. Allen, M. J. Bowick and A. Lahiri, Mod. Phys. Lett. {\bf A6}, 559 (1991).

\vspace{1mm}

A. Lahiri, Mod. Phys. Lett. {\bf A8}, 2403 (1993). 

\vspace{1mm}

H. Sawayanagi, Mod. Phys. Lett. {\bf A10}, 813 (1995). 

\vspace{1mm}

R. Amorim and J. Barcelos-Neto, Mod. Phys. Lett. {\bf A10}, 917 (1995). 

\vspace{1mm}

J. Barcelos-Neto and A. Cabo, Z. Phys. {\bf C74}, 731 (1997). 
%
%[8]
\item D. Z. Freedman and P. K. Townsend, Nucl. Phys. {\bf B177}, 282 (1981). 

\vspace{1mm}

J. Barcelos-Neto, A. Cabo and M. B. D. Silva, Z. Phys. {\bf C72}, 345 (1996). 

\vspace{1mm}

D. S. Hwang and C. Y. Lee, J. Math. Phys. {\bf 38}, 30 (1997).  

\vspace{1mm}

A. Lahiri, Phys. Rev. {\bf D55}, 5045 (1997).  

\vspace{1mm}

J. Barcelos-Neto and  S. Rabello, Z. Phys. {\bf C74}, 715 (1997). 
%
%[9]
\item E. Bergshoeff, M. de Roo, B. de Wit and P. van Nieuwenhuizen, 
Nucl. Phys. {\bf B195}, 97 (1982).  
%
%[10]
\item M. B. Green, J. H. Schwarz and E. Witten, {\it Superstring Theory} 
Vol. 2 (Cambridge University Press, New York, 1987)  
%
%[11]
\item T. Kimura, Prog. Theor. Phys. {\bf 64}, 357 (1980). 

\vspace{1mm}

H. Hata, T. Kugo and N. Ohta, Nucl. Phys. {\bf B178}, 527 (1981). 

\vspace{1mm}

J. Thierry-Mieg and L. Baulieu, Nucl. Phys. {\bf B228}, 259 (1983). 

\vspace{1mm}

M. Henneaux and C. Teitelboim, {\it Quantization of Gauge Systems} 
(Princeton University Press, New Jersey, 1992).
%
%[12]
\item R. A. Bertlmann, {\it Anomalies in Quantum Field Theory} 
(Oxford University Press, New York, 1996). 
%
%[13]
\item G. F. Chapline and N. S. Manton, Phys. Lett. {\bf B120}, 105 (1983). 
%
%[14]
\item S. Deguchi and T. Nakajima, Mod. Phys. Lett. {\bf A12}, 111 (1997). 
%
%[15]
\item L. Baulieu, Phys. Lett. {\bf B126}, 455 (1983);  
Nucl. Phys. {\bf B227}, 157 (1983). 
%
%[16]
\item H. Georgi and S. L. Glashow, Phys. Rev. {\bf D6}, 429 (1972).  
%
%[17]
\item A. N. Redlich, Phys. Rev. Lett. {\bf 52}, 18 (1984); Phys. Rev.  
{\bf D29}, 2366 (1984). 
%
%[18] 
\item T. Fujiwara and S. Kitakado, Mod. Phys. Lett. {\bf A8}, 1639 (1993).

\end{enumerate}

\end{document}